\documentclass{emulateapj}

\shorttitle{Rapid orbital decay caused by circumbinary disks}
\shortauthors{Chen \& Podsiadlowski}

\begin{document}


\title{Rapid orbital decay in detached binaries: evidence for circumbinary disks }


\author{Wen-Cong Chen$^{1,2,3}$, and Philipp Podsiadlowski$^{2,3}$ }
\affil{$^1$ School of Physics and Electrical Information, Shangqiu Normal University, Shangqiu 476000, China;\\
 $^2$ Department of Physics, University of Oxford, Oxford OX1 3RH, UK;\\
 $^3$ Argelander-Insitut f\"{u}r Astronomie, Universit\"{a}t Bonn, Auf dem H\"{u}gel 71, 53121 Bonn, Germany;
chenwc@pku.edu.cn
}



\begin{abstract}
  Some short-period, detached binary systems have recently been
  reported to experience very rapid orbital decay, much faster than is
  expected from the angular-momentum loss caused by gravitational
  radiation alone. As these systems contain fully convective stars,
  magnetic braking is not believed to be operative, making the large
  orbital-period derivative puzzling.  Here we explore whether a
  resonant interaction between the binary and a surrounding
  circumbinary (CB) disk could account for the observed orbital
  decay. Our calculations indicate that the observed orbital-period
  derivatives in seven detached binaries can be produced by the
  resonant interaction between the binary and a CB disk if the latter
  has a mass in the range of $10^{-4}-10^{-2}~ M_{\odot}$, which is of
  the same order as the inferred disk mass ($\sim2.4\times 10^{-4}~
  M_{\odot}$) in the post-common-envelope binary NN Ser.
\end{abstract}


\section{Introduction}
Interacting binaries are generally systems in which one or both
components overflow their Roche lobe(s); they are ideal probes for
testing stellar and binary evolutionary theory. During the evolution
of interacting binaries, orbital angular-momentum loss mechanisms play
a vital role in driving mass transfer and the orbital evolution. The
most important of these are due to gravitational radiation, mass loss,
and magnetic braking. However, it is often difficult in interacting
binaries to distangle the effects of angular-momentum loss from the
effects of mass transfer.

In contrast, in detached binaries there is usually no significant mass
transfer; this makes it possible -- in principle -- to constrain any
systemic angular momentum loss directly from an orbital period change.
Recently, some detached binaries have been found to experience very
rapid orbital decay.  The detached eclipsing binary SDSS
J14-3547.87+373338.5 (hereafter J1435) with an orbital period of
0.126\,d has a white dwarf primary \citep{eise06} and a companion star
of spectral type M4--M6 \citep{stei08}, which is most likely fully
convective. \cite{qian16} reported that the orbital period of this
system decreases continuously at a rate $\dot{P}=-8.04\times
10^{-11}~\rm s\,s^{-1}$.  Generally, orbital period changes in binary
systems may be caused by three mechanisms: (1) genuine orbital
angular-momentum loss, (2) the so-called Applegate mechanism
(Applegate 1992), and (3) the existence of a third body in the system.
Gravitational radiation alone only produces a period derivative of
$\dot{P}_{\rm gr}=-8.1\times 10^{-14}~\rm s\,s^{-1}$, which is three
orders of magnitude smaller than the observed value. Meanwhile, in the
standard theory a fully convective star does not produce magnetic
braking \citep{verb81,rapp83,spru83}, a key ingredient in explaining the
period gap of cataclysmic variables.

In Table 1, we summarize the relevant parameters for seven detached
binary systems with an observed orbital-period derivative. For a
detached binary, the orbital-period derivative induced by
gravitational radiation is given by
\begin{equation}
\dot{P}_{\rm gr}=-\frac{96G^{3}}{5c^{5}}\frac{M_{1}M_{2}(M_{1}+M_{2})}{a^{4}}P,
\end{equation}
where $G$ is the gravitational constant, $c$ the speed of light in
vacuo, $a$ the orbital separation, $P$ the orbital period, and $M_{1}$
and $M_{2}$ the masses of the two components, respectively. As the
table shows, the orbital-period derivatives that can be caused by
gravitational radiation alone are $1-5$ orders of magnitude smaller
than the values observed.

If one assume that the secondary ($M_{2}$) experiences magnetic braking,
the standard model for magnetic braking given by \cite{rapp83} predicts an
orbital-period derivative
\begin{eqnarray}
\dot{P}_{\rm mb}=-1.4\times 10^{-12}\left(\frac{ M_{\odot}}{M_{1}}\right)\left(\frac{M_{1}+M_{2}}{ M_{\odot}}\right)^{1/3}\nonumber \\
\left(\frac{R_{2}}{ R_{\odot}}\right)^{\gamma}\left(\frac{\rm d}{P}\right)^{7/3}~\rm s\,s^{-1}.
\end{eqnarray}
Even if magnetic braking were not to cease completely for stars without
radiative core \citep{king02,andr03,pret08,qian15}, this mechanism
(with the standard value of $\gamma=4$) could not be responsible for
the rapid orbital decay in these seven detached binaries (see Table 1).

If the observed period decrease is just a short-term phenomenon
(rather than a secular decrease), it could possibly arise from the
Applegate mechanism \citep{appl92}. It is generally thought that the
stellar magnetic field results from magnetic dynamo action. Here the
interplay between differential rotation and cyclonic convection
induces a magnetic activity cycle as follows: the shear due to the
differential rotation drives a poloidal field to form a toroidal
subsurface field; due to the convection, the toroidal field moves to
the star's surface and twists it; finally, the Coriolis force converts
this twist into a poloidal field, thus finishing the cycle
\citep{appl92}. In this picture, the magnetic activity induces
angular-momentum transfer between different convective zones of the
cool star, changing the rotational velocity of the star, and hence,
due to the tidal locking, the orbital period of the binary.  However,
the energy necessary to make this mechanism operate cannot be provided
by the total radiation energy in the case of NN Ser \citep{brin06} and
V2051 Oph \citep{qian15}. Therefore, it is still a puzzle why these
detached binaries experience such rapid orbital shrinking. The
solution of this problem may offer some more general, valuable clues
for the theory of single and binary stars.

Assuming that the orbital-period changes in these seven detached
binaries originate from genuine angular-momentum loss from the binary,
we attempt in this Letter to diagnose whether a circumbinary (CB) disk
around these systems could provide the solution by causing the
transfer of orbital angular momentum from the binary to the disk.  In
section 2, we describe the CB disk model and compare the predicted
orbital-period derivatives with the observed values. In Section 3, we
summarize the results with brief conclusions and a discussion.

\section{circumbinary disk model}

The existence of CB disks has previously been deduced from the
analysis of the blackbody spectrum in the black-hole X-ray binaries
A0620$-$00 and XTE J1118+480. \cite{muno06} found that the inferred
emitting surface areas of the observed excess mid-infrared emission
are approximately two times larger than the binary orbit. Subsequent
observations with the Wide-Field Infrared Survey Explorer have
indicated that these two sources may hide CB disks
\citep{wang14}. Recently, ALMA detection of flux at 1.3 mm for the
post-common-envelope binary NN Ser has been proposed to likely
originate from the thermal emission from a dust disk with a mass
of $2.4\pm0.6\times 10^{-6}~ M_{\odot}$ \citep{hard16}. These
observations provide strong evidence that CB disks may exist around
some detached binaries. So far, the process by which such CB disks
form has not yet been fully understood. CB disks may be produced by
some common-envelope material that has not been ejected from the
system for a binary like NN Ser. Alternatively, the origin of a CB
disk may be closely related to the mass outflow processes during mass
transfer. For example, it has been argued in the case of postulated CB
disks around millisecond pulsar binaries that, during hydrogen-shell
flashes in the helium secondary, the transferred material exceeding
the Eddington accretion rate may be expelled through L2, causing the
formation of a CB disk \citep{anto14}.

Because of the relatively large orbital angular momentum, mass loss
during mass transfer may form a disk surrounding the binary system
rather than leave it \citep{heuv73,heuv94}. Assuming a small fraction
of the transferred mass feeds a CB disk, \cite{spru01} and
\cite{taam01} found that the torque provided by the CB disk could
produce a range of variation of mass-transfer rates that is compatible
with the range of mass-transfer rates seen in cataclysmic
variables. Other relevant studies show that the existence of a CB disk
can provide an efficient angular-momentum loss mechanism, which
enhances the mass-transfer rate and causes the secular shrinking of
the binary orbit \citep{chen06a,chen06b,chen07,chen15}.

\begin{table*}
\centering
\begin{minipage}{170mm}
\caption{Key parameters for seven detached binary systems with
  observed orbital-period derivatives. The columns list (in order):
  the source name, type of primary, primary mass, secondary mass,
  secondary radius, orbital period, orbital separation, observed
  orbital period derivative,  orbital period
  derivative induced by gravitational radiation, orbital-period
  derivative induced by magnetic braking, and references.}
\begin{tabular}{lllllllllll}
  \hline\hline\noalign{\smallskip}
Sources & Primary types &$M_{1}$ & $M_{2}$   & $R_{2}$ & $P$ &  $a$ & $\dot{P}$  & $\dot{P}_{\rm gr}$&$\dot{P}_{\rm mb}^{\ast}$ &References\\
        &  &($ M_{\odot})$ & ($ M_{\odot}$)& ($ R_{\odot}$) & (days) & ($ R_{\odot}$)  &($10^{-11}\rm s\,s^{-1}$)  & ($10^{-13}\rm s\,s^{-1}$)& ($10^{-13}\rm s\,s^{-1}$)& \\
 \hline\noalign{\smallskip}
J1435      & white dwarf &0.5 &0.21&0.23  &0.126 &0.94 & $-8.04$ & $-0.81$&-8.8 & 1,2,3\\
V2051 Oph  & white dwarf &0.78&0.15&0.15  &0.0625&0.65 &$-0.162$ &-2.6& -5.7 & 4,5  \\
NN Ser     & white dwarf &0.535&0.111&0.18&0.13  &0.93  &$-0.60$ &-0.45&-2.8  & 6,7,8  \\
HU Aqr     & white dwarf &0.88&0.20 & 0.22&0.087&0.85  &$-0.56$  & -2.2 &-11.4  &9,10  \\
NY Vir     & subdwarf B  &0.46&0.14 & 0.18&0.101&0.77  &$-0.92$  & -0.76 &-5.7 & 11,12   \\
HW Vir     & subdwarf B  &0.48&0.14 & 0.18&0.117&0.86  &$-2.12$  &-0.61&-3.9  & 13,14   \\
WY Cancri  & G5          &0.81&0.31 & 0.58&0.829&3.85  &$-33.0$  &-0.07 &-3.1 & 15,16  \\
\noalign{\smallskip}\hline
\end{tabular}
\end{minipage}
\\ \textbf{References}. (1) \cite{stei08}; (2) \cite{pyrz09}; (3) \cite{qian16}; (4) \cite{bapt07}; (5) \cite{qian15};
(6) \cite{haef89}; (7) \cite{brin06}; (8) \cite{qian09}; (9) \cite{schw09}; (10) \cite{qian11};
(11) \cite{kilk98}; (12) \cite{qian12}; (13) \cite{wood99}; (14) \cite{qian08}; (15) \cite{zeil90}; (16) \cite{heck98}.
\\$\ast$ \textbf{Note}.  We only consider the magnetic braking contribution of secondary stars.
\end{table*}

We now assume that the seven detached binaries with rapid orbital
decay are surrounded by CB disks. Recent MHD simulations show
  that, for the mass ratios applicable to our situations, both
  components of the binary are located inside the inner edge of the
  CB-disk \citep{gunt02,shi12,shi15}. Therefore, we assume that there
  is a wide gap between the binary systems and the disk, and we
  neglect any mass inflow from the CB disk into the inner region (but
  also see the further discussion in section 3).  Because of Keplerian
  rotation, CB disks have a lower angular velocity than the binary
  systems. By the tidal interaction, the angular momentum is
  continuously transferred from the binary to the inner edge of the
  disk. Subsequently, the angular momentum is transported from the
  inner edge to the outer region of the disk by the waves and internal
  viscous stresses \citep{arty94}. As a result, the disk rapidly
  spreads outwards to redistribute the angular momentum. The rate by
which the orbital separation decays due to the resonant interaction
between the binary and the CB disk is given by \citep{lubo96,derm13} as
\begin{equation}
\frac{\dot{a}}{a}=-\frac{2l}{m}\frac{M_{\rm d}\alpha}{\mu}\left(\frac{H}{R}\right)^{2}\frac{a}{R}\Omega,
\end{equation}
where $l$ and $m$ are the time-harmonic number and the azimuthal
number (Artymowicz \& Lubow 1994); $M_{\rm d}$ and $\mu$ are the CB
disk mass and the reduced mass of the binary, respectively; $\alpha$
is the viscous parameter, $\Omega$ the orbital angular velocity, $H$
and $R$ the thickness and the half-angular-momentum radius of the
disk, respectively.

For a detached binary, Kepler's third law
($G(M_{1}+M_{2})/a^{3}=4\pi^{2}/P^{2}$) leads to
\begin{equation}
\frac{\dot{P}}{P}=\frac{3}{2}\frac{\dot{a}}{a}.
\end{equation}
The objects we explore in this study are all detached binaries with
short orbital periods. We therefore assume that their eccentricities
are small due to the tidal interactions. When the eccentricity $e\leq
0.1\sqrt{\alpha}$, the resonances are very weak to drive the eccentricity, and $m=l$
\citep{derm13} \footnote{For larger eccentricities
$0.1\sqrt{\alpha}\la e\la 0.2$, the $m=2, l=1$ resonance is the
strongest contribution driving the eccentricity.}. Combining equations (1) and (2), the predicted orbital
period derivative can be written as
\begin{equation}
\dot{P}=-6\pi\frac{M_{\rm d}\alpha}{R}\left(\frac{H}{R}\right)^{2}\frac{a}{\mu}.
\end{equation}
Equation (5) shows that the orbital-period derivative is determined by
two factors: a degenerate factor that depends on a combination of the
CB disk parameters (${(M_{\rm d}\alpha H^{2})}/{R^{3}}$) and a
parameter that depends on the properties of the binary
($a/\mu$). Based on the measured CB disk mass in NN Ser, we can
constrain the first factor and hence the CB disk parameters (i.e.,
$\alpha, H,$ and $R$), and then use his calibration for the other six
sources.

If the gas-to-dust ratio of a CB disk is similar to that of the
interstellar medium ($\sim 100$, Muno \& Mauerhan 2006), the total
mass of the CB disk in NN Ser can be estimated to be $2.4\times
10^{-4}~\rm M_{\odot}$. Assuming that the orbital decay of NN Ser
arises from the tidal torque of a CB disk, we obtain ${\alpha
  H^{2}}/{R^{3}}=1.9\times 10^{-21}~\rm cm^{-1}$ from equation (5).
In this work, we adopt a relatively thin CB disk and low viscous
parameter, i.e.\ $H/R=0.03$, $\alpha=0.01$, which yields $R=315~\rm
AU$.  The inner radius ($r_{\rm in}$) of the CB disk relies on the
resonant torque between the binary and the CB disk, which pushes the
disk outwards. Based on SPH simulations for the inner radius of the CB
disk at various eccentricities \citep{arty94}, \cite{derm13} fitted a
simple formula:
\begin{equation}
r_{\rm in}(e,\Re)=1.7+0.375\,{\rm log}(\Re \sqrt{e})~\rm AU,
\end{equation}
where $\Re=(H/R)^{-2}\alpha^{-1}\approx 10^{5}$ is the Reynolds number
of the CB disk gas.  This implies that the inner radius of the CB disk
is not sensitive to the eccentricity ($e=10^{-4}$ leads to $r_{\rm
  in}=2.825~\rm AU$, whereas $e=10^{-6}$ yields $r_{\rm in}=2.45~\rm
AU$). In this work, we take $e=10^{-4}$ and the outer radius of the
disk $r_{\rm out}=1000~\rm AU$, so that $R=(r_{\rm in}+r_{\rm
  out})/4+\sqrt{r_{\rm in}r_{\rm out}}/2\approx300~\rm AU$
\footnote{We assume that the relation between the surface density and
  the distance from the disk center obeys $\sigma(r)\propto
  r^{-2}$. }, which is in good agreement with the value of $R$ derived
for NN Ser.

  To examine whether a CB disk could be responsible for the orbital
  decay of our detached binaries, we show in Figure 1 a comparison of
  the predicted orbital-period derivatives in the CB disk model with
  observations in a $\dot{P}-a/\mu$ diagram. All theoretical curves
  (using equation~3) are based on the same CB disk parameter $
  {\alpha H^{2}}/{R^{3}}$ as deduced for NN Ser. The estimated CB disk mass
  in NN Ser ($M_{\rm d}=2.4\times10^{-4}~ M_{\odot}$) is shown as a
  solid line. Since one would generally expect a range of CB disk
  masses for different systems, we also show $\dot{P}$ as a function
  of $a/\mu$ for disk masses varying by two orders of magnitude:
  $10^{-2}$ (dashed line), $10^{-3}$ (dotted line), and $10^{-4}~
  M_{\odot}$ (dashed-dotted line). Five of the detached binaries are
  located within a fairly narrow range of CB disk masses of
  $10^{-4}-10^{-3}~ M_{\odot}$, while both SDSS J143547.87+373338.5
  and WY Cancri would require a rather heavy CB disk around $10^{-2}~
  M_{\odot}$.  This suggests that the CB disk model can in principle
  explain the observations for a range of assumed disk masses and be
  responsible for the observed orbital-period derivatives.
\begin{figure}
\centering
\includegraphics[width=1.15\linewidth,trim={0 0 0 0},clip]{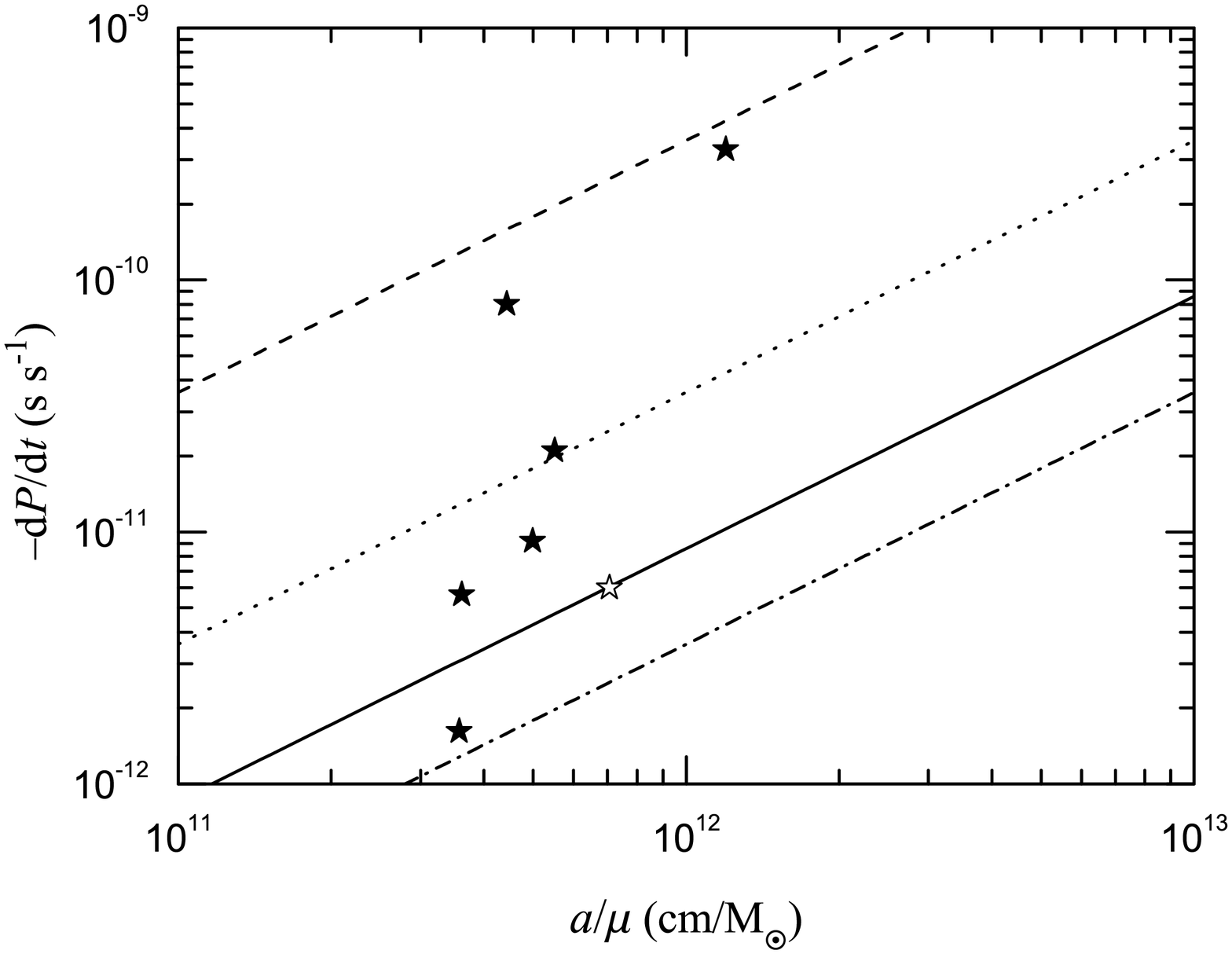}
\caption{Comparison of the predicted orbital-period derivatives in the
  CB disk scenario with observations in the $\dot{P}-a/\mu$
  diagram. The solid stars denote six detached binaries, the open star
  denotes NN Ser. The solid, dashed, dotted, and dashed-dotted curves represent
  CB disk masses of $2.4\times10^{-4}$ (the inferred CB disk mass for NN
  Ser), $10^{-2}$, $10^{-3}$, and $10^{-4}~ M_{\odot}$,
  respectively. } \label{fig:orbmass}
\end{figure}

The dynamical interaction between the binary and the CB disk can
induce an eccentricity in a binary system \citep{arty94}. However, the tidal
force in a secondary star with a relatively deep convective envelope will try
to circularize the orbit. The circularization
timescale for a binary containing a $0.5~ M_{\odot}$ main-sequence
star and a $0.8~ M_{\odot}$ white dwarf can be estimated as \citep{hurl02}
\begin{equation}
\tau_{\rm cir}\approx 60 \left(\frac{a}{R_{\odot}}\right)^{8}~\rm yr.
\end{equation}
For a small eccentricity, $\dot{e}=-50e\dot{a}/(\alpha a)$
\citep{lubo96}, the timescale for driving eccentricity can be written as
\begin{equation}
\tau_{\rm ecc}\approx\frac{e}{\dot{e}}=-\frac{\alpha a}{50\dot{a}}.
\end{equation}
With the exception of WY Cancri, all other sources have a short
circularization timescale, $\tau_{\rm cir}\la40~\rm yr$. According to
equation (8), the eccentricity growth timescale is inversely
proportional to $\dot{P}$. SDSS J143547.87+373338.5 has the shortest
eccentricity growth timescale of $\tau_{\rm ecc}\approx 1.3\times
10^{4}~\rm yr$. Since $\tau_{\rm cir}\ll \tau_{\rm ecc}$, the tidal
force of the secondary is sufficiently strong to keep the orbit
approximately circular.

\section{Discussion and Summary}
In a previous study, we have already investigated whether a CB disk
could be responsible for the large orbital-period derivative detected
in NN Ser \citep{chen09} and WY Cancri \citep{chen13}. We assumed that
a fraction of the wind loss from the donor star feeds the CB disk,
which would yield a tidal torque by the interaction between the disk
and the binary to extract orbital angular momentum. To produce the
observed orbital-period derivatives, the donor stars would have to
have a very high wind mass-loss rate and a relatively high fraction of
the wind feeding the CB disk. Therefore, the possibility that the CB
disk fed by a wind alone to produce the rapid orbital decay observed
in these two sources could be ruled out \citep{chen09,chen13}.

In this Letter, we employed a very different CB disk model. Six of the
seven detached binaries listed in Table 1 have ultra-compact orbits
(all sources except for WY Cancri have an orbital period less than 0.2
d), which are all likely to be post-common-envelope binaries. In the
final stage of a common-envelope phase, a fraction of the orbital
energy of the binary is believed to expel most of the envelope, giving
rise to the birth of a compact binary \citep{ivan13}. However, not all
the CE material needs to be ejected, and the remaining material may
collapse into a disk surrounding the binary, i.e. form a CB disk
\citep{spru01} around the newly formed more compact, detachted binary.
Our estimates show that the existence of a CB disk can produce the
observed orbital-period derivatives in such systems. The interaction
torque between the binary and the CB disk plays a vital role in the
shrinking of the binary orbit. Based on the disk parameters derived by
the observed orbital-period derivative and the inferred mass
($\sim2.4\times10^{-4}~ M_{\odot}$) of the candidate CB disk in NN
Ser, the disk mass in seven sources could be constrained to be
$10^{-4}-10^{-2}~ M_{\odot}$, which is close to the observed disk mass
range \citep[$10^{-4}-10^{-3}~ M_{\odot}$; ][]{giel07}. The
  viscous timescale at the inner edge of the disk, $\tau=
  \alpha^{-1}(H/R)^{-2}(r_{\rm in}/a)^{3/2}\Omega^{-1}$, can be
  estimated to be $\sim 10^{5}~\rm yr$ for NN Ser (adopting the same
  parameters as in section 2).  \cite{haef04} obtained a cooling time
  $1.3~\rm Myr$ for the white dwarf in NN Ser.  This implies that
  viscous torques operate sufficiently fast that they can successfully
  transport the angular momentum to the outer region of the disk, but
  also that the phase of fast orbital decay (at least several viscous
  time scales) is long enough to be observable in a system like NN
  Ser.

Our scenario neglected any gas flow in or across the gap
between the disk and the binary. On the other hand, detailed
simulations indicate that some gas at the inner boundary of the disk
  may penetrate the gap and be accreted onto one or both binary
  components \citep{gunt02,shi12,shi15}. Assuming the inflowing gas
  possess the specific angular momentum ($j=\sqrt{G(M_{1}+M_{2})r_{\rm
      in}}$) at the inner edge of the CB disk, the rate of
  orbital-period increase of the binary for an accretion rate
  $\dot{M}$ can be estimated to be
  \begin{equation}
  \frac{\dot{P}}{P}\approx
  3\frac{\dot{J}_{\rm
      acc}}{J}=\frac{3j\dot{M}}{\mu\sqrt{G(M_{1}+M_{2})a}}=\frac{3\dot{M}}{\mu}\left(\frac{r_{\rm
      in}}{a}\right)^{1/2},
  \end{equation}
  where $\dot{J}_{\rm acc}$ is the
  angular-momentum-increasing rate by the accretion, $J$ is the total
  angular momentum of the binary.  For NN Ser, the gas inflow from
  the disk would not change the tendency of orbital decay if the
  accretion rate $\dot{M}<2.5\times10^{-11}~M_{\odot}\,\rm
  yr^{-1}$. This threshold accretion rate is seven orders of magnitude
  higher than the inferred value
  ($\sim9.0\times10^{-18}~M_{\odot}\,\rm yr^{-1}$) for a CB disk
  surrounding a white dwarf binary \citep{fari17}. Therefore, it seems
  that the gas inflow of the CB disk cannot significantly alter the
  evolution of the binary in our scenario.

One may expect that the mass contained in a CB disk would evolve
because of mass loss via a disk wind.  However, it is difficult to
estimate the wind loss rate at the moment.  In principle, the change
of the orbital period derivative could be used to obtain an indirect
estimate of the disk wind loss rate. According to equation (5), the CB
disk mass changes according to $\bigtriangleup M_{\rm d}\propto
\bigtriangleup \dot{P}$, where $\bigtriangleup \dot{P}$ is the change
of the orbital period derivative. V2051 Oph has the smallest estimated
CB disk mass ($\sim 10^{-4}~ M_{\odot}$) and therefore provides the
best candidate for testing the CB disk scenario.  If $\dot{P}$
decreases by a factor of 1\% over a period of 1000 yr, the CB
disk mass would accordingly decrease by 1\%, producing a disk mass loss rate
of $\sim 10^{-9}~ M_{\odot}\,\rm yr^{-1}$.

In principle, our CB scenario strongly depends on the migration rate. \cite{kocs12} simulated
the co-evolution of binaries and CB disks, and found that, gas would pile up outside the binary's
orbit rather than creating a cavity for some system with smaller separations and masses. In this case,
the disk properties are intermediate between between the weakly perturbed case (Type I migration)
and the case with a gap (Type II migration). As a result, the migration rate of the secondary
in this 'Type 1.5' regime is obviously slower than both Type I and
Type II rates. Therefore, the CB disk would not be responsible for the rapid orbital decay of these detached
binaries if the migration rates are similar to Type 1.5 rate. In addition, we neglect the persistent co-rotation
torques, which should tend to drive an outwards migration.

Besides the Applegate mechanism, a third body can also produce cyclic
period changes in close binaries. For example, \cite{qian09} proposed
that a tertiary component with a mass of $11.1~\rm M_{\rm Jupiter}$
orbiting NN Ser at an orbital radius of $d_{3}=3.29~\rm AU$ could give
rise to the observed orbital period derivative. However, this
variation would be periodic not secular, and it should therefore be
possible to distinguish this from genuine angular-momentum loss from
the binary.  Hence the observed orbital decays may provide indirect
evidence for the existence of CB disks in these systems.  Long-term
observations should further allow to strengthen this conclusion.  In
addition, CB disks may be observable with L band ($3-4~\rm\mu m $)
observations due to the continuum contribution of dust emission
\citep{spru01}. Therefore, future detailed multi-waveband observations
for these detached binaries are required to confirm or refute our
scenario.

\acknowledgments {We are grateful to the anonymous referee for helpful comments
improving the manuscript. We also thank S.-B. Qian for providing us the relevant
  binary parameters of some detached binaries. This work was partly
  supported by the National Natural Science Foundation of China (under grant
  number 11573016), the Program for Innovative Research Team (in
  Science and Technology) at the University of Henan Province, and the
  China Scholarship Council. This work has also been supported by a
  Humboldt Research Award to PhP at the University of Bonn.}

\end{document}